\documentstyle[10pt,fleqn,epsfig,multicol,amssymb,amsmath]{article}

\makeatletter

\newenvironment{figurehere}
  {\def\@captype{figure}}
  {}
\makeatother

\begin{document}

\begin{center}

 {\large{\bf
 NEURAL NETWORK ANALYSIS OF THE MAGNETIZATION REVERSAL
 IN MAGNETIC DOT ARRAYS}}

 \vspace{0.9cm}

    { Martin GMITRA and Denis HORV\'ATH } \\
    {\small
      Department of Theoretical Physics and Astrophysics, Faculty of Natural Science \\
      University of P.J.\v{S}af\'arik, Moyzesova 16, 040 01 Ko\v{s}ice, tel. 055/622 22 21,\\
      E-mails: gmitra@host.sk, horvath@kosice.upjs.sk\\
    }

\vspace{0.9cm}

{\small\bf SUMMARY}\\
\end{center}

\vspace{-0.3cm}

{\small\em
We simulated the remagnetization dynamics of the ultra-dense
and ultra-thin magnetic dot array system with dipole-dipole
and exchange coupling interactions. Within the proposed
2D XY superlattice model, the square dots are modeled
by the spatially modulated exchange-couplings.
The dipole-dipole interactions were approximated by
the hierarchical sums and dynamics was reduced
to damping term of the Landau-Lifshitz-Gilbert equation.
The simulation of $40 000$ spin system leads to nonequilibrium
nonuniform configurations with soliton-antisoliton pairs
detected at intra-dot and inter-dot scales. 
The classification of intra-dot magnetic configurations
was performed using the self-adaptive neural networks
with varying number of neurons.
}

\vspace{0.5cm}
\hspace{-0.5cm}$\mbox{\boldmath $Keywords:$}$ {\small\em dot array, neural network model, XY model, numerical simulation}
\vspace{0.3cm}

\begin{multicols}{2}

\vspace{0.5cm}
\hspace{-0.5cm}{\bf{1. INTRODUCTION}}
\vspace{0.4cm}

Many interesting applications of the periodic magnetic
dot arrays are expected in the field of the magnetic recording
{\cite{Prinz95}} and magnetic sensors. The typical properties
brought by the array geometry is the limited length of the spin chains
coupled by the exchange coupling and the uniqueness of
the magnetostatic interactions at the scales comparable
or larger than dot size.

The powerful tools to study the static and dynamic properties of magnetic nanostructures
including the thin films and isolated small particles represent micromagnetic {\cite{Liou2001}},
and Monte Carlo \cite{Levy1999,Iglesias2000,Kachkachi2000,Lieneweber1999}
simulations. They are focussed to the analysis of the nonuniform magnetization states.
On a purely  qualitative
level the intra-dot nonuniformities can be characterized
as a mixtures of vortices,
flowers, domain walls, etc.. Our recent activity is related
to the search for the tools
allowing the systematic classification
of the simulated magnetic structures.

The methodology presented in this paper
is inspired by the progress in the theory of artificial
neural networks, that are nonlinear models suitable to
reduce, classify or interpolate the information involved
in the systems of the complex patterns {\cite{Haykin98}}.
In the previous paper {\cite{Horvath2001}}
we suggested the implementation of radial basis function networks
to model the magnetic dot array magnetization processes.
The lack of this proposition was the absence of
the network learning which needs the
data support from physically acceptable
micromodel simulations.

The above reasons led us to make some progress
towards the opposite direction by using the neural
network analysis of the configurations generated
by the remagnetization processes described
by the XY superlattice model of the ultra-dense array
of the square dots limited to in-plane spin polarization.
The motivation for the study of ultra-dense spin
structures is the special interest for the interfacial
inter-dot effects. 
The experimental studies 
of the magnetization reversal are generally not
in agreement with the familiar
model of coherent rotation. Instead, complex
processes such as curling and buckling are observed.
The model structures were obtained by
simulations based on the numerical
solving of Landau-Lifshitz-Gilbert equation.
The simulations bring
the usual problem of effective
data post-processing \cite{Schabes87}.
In the present paper the data 
were treated
by means of so called {\em adaptive re\-sonance neural network}
ART, which is introduced in section~3.
This tool was used to classify
the intra-dot magnetic 
configurations, which are the transients
of the magnetization reversal process.
The ART choice is connected with its ability
to vary the number of neurons
(network topology) according 
to complexity of treated inputs,
which are in our case activated by the intra-dot magnetic
configurations. In the section~4 we concentrate
to the analysis of data provided by the simu\-lated
magnetization reversal.

\vspace{0.5cm}
\hspace{-0.5cm}{\bf{2. XY MODEL OF ULTRA-DENSE \\ \hspace*{0.5cm}ARRAY}}
\vspace{0.5cm}

The details of the remagnetization dynamics
are studied by means of the model of finite magnetic superlattice
system described by the XY spin Hamiltonian
\begin{multline}
\label{hamiltonian}
{\cal H} = -\frac{K}{2} \sum_{{{\alpha,\beta}\atop{\xi,\zeta=x,y}}} S_\alpha^\xi D_{\alpha\beta}^{\xi\zeta} S_\beta^\zeta
 \\ -\frac{J}{2}\sum_{\alpha,\beta} \epsilon_{\alpha,\beta} {\bf S}_\alpha \cdot {\bf S}_\beta
 - H\sum_\alpha {\bf S}_\alpha\cdot{\bf e}_x
\end{multline}
The first sum of Eq.(\ref{hamiltonian})
represents the dipolar contribution
controlled by the constant $K$,
where ${\bf S}_\alpha$ ($|{\bf S}_{\alpha}|=1$)
is the planar spin variable and $D_{\alpha\beta}^{\xi\zeta}$
is the demagnetization tensor of superlattice;
$\alpha,\beta$ are site indices.
The second sum represents the exchange energy contribution
with periodically modulated nearest-neighbor exchange
coupling interaction $J\epsilon_{\alpha\beta}$, where
$\epsilon_{\alpha\beta}\in\{0,1\}$ (see Fig.~\ref{schema_J}).
The third sum represents standard Zeeman term
where ${\bf e}_x$ is the Cartesian unit vector oriented along
the main axis of superlattice.
Presented model belongs to the idealized structure,
where the patterning attains the maximum geometric filling
independent of dot size ($L_0$ linear size of dot).
\begin{figurehere}
\begin{center}
\epsfig{file=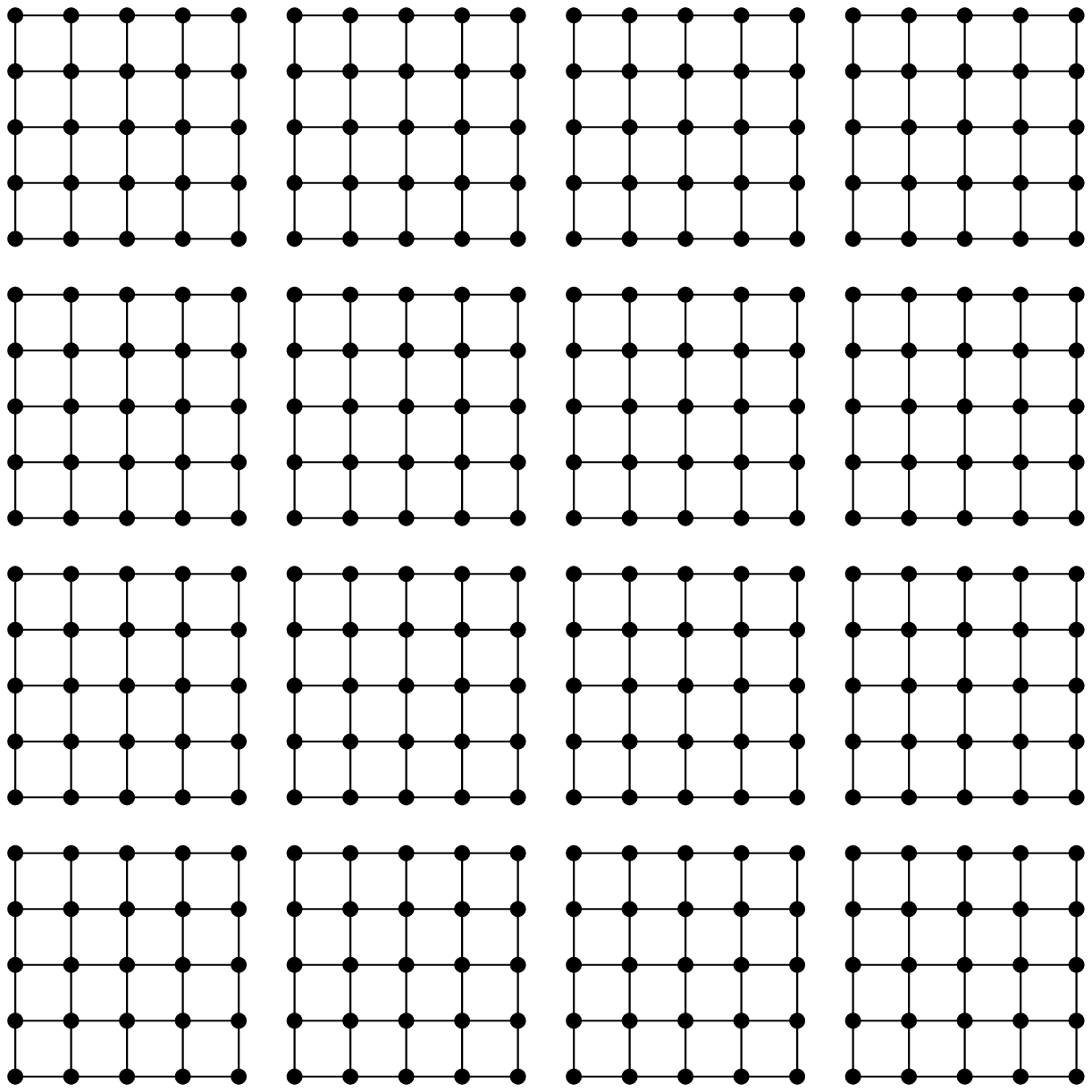,width=6.0cm,angle=0.0}
\caption{\small
 \label{schema_J}
 The schematic plot of the exchange coupling
 (bonds represents $\epsilon_{\alpha\beta}=1$) modulation.
 The system contains 16 dots, each of them consists of 25 spins.}
 \end{center}
\end{figurehere}

The model dynamics was described by the Landau-Lifshitz-Gilbert
equation, that in 2D case reduces to relaxation term.
The numerical integration scheme was derived
from the recursive propagator form
\begin{equation}
\label{intscheme}
{\bf S}_\alpha(t+\Delta t) =
\hat{\bf U}\left(
\int_t^{t+\Delta t}
{\omega}_\alpha(\tau)
\,
\mbox{d} \tau
\,
\right)
\,{\bf S}_\alpha(t)\,,
\end{equation}
where ${\omega}_\alpha(t)
=(   {\bf S}_\alpha(t) \cdot {\bf e}_x              )
 (   {\bf h}_\alpha(t)\cdot {\bf e}_y   ) -
 (   {\bf S}_\alpha (t) \cdot {\bf e}_y             )
 (   {\bf h}_\alpha(t)\cdot {\bf e}_x   ) $
is the instant angular frequency of the spin rotation,
${\bf h}_\alpha(t)=-\delta{\cal H}/\delta S_\alpha$
is the local effective field; $t$ is the time,
$\Delta t$ is the integration step.
The unitarity of the spin rotation matrix
\begin{equation}
\label{rot_matrix}
 \hat{\bf U}(\varphi) = \left(
                     \begin{array}{lr}
                      \cos\varphi & -\sin\varphi \\
                      \sin\varphi & \cos\varphi
                     \end{array}
                     \right)
\end{equation}
implies that numerical integration scheme conserves the spin vector
size independently of additional approximation of
integral in Eq.(\ref{intscheme}).
The computationally
demanding task of the integration
is the calculation of the dipolar fields,
where essential reduction brings the hierarchical
summation \cite{Miles91}. In the case of array,
the most natural hierarchy level choice is the block
association with the single intra-dot moment.

\vspace{0.5cm}
\hspace{-0.5cm}{\bf{3. THE ART CLASSIFICATION OF\\ \hspace*{0.5cm}INTRA-DOT CONFIGURATIONS}}
\vspace{0.5cm}

 The artificial neural network models are inspired
 by the physiology, and mimic the neurons and synaptic
 connections of a brain. They can be considered as mappings
 \cite{Haykin98} constructed from given activation functions.
 The unknown parameters of these functions called synaptic weights
 are adjusted by training. The fascinating feature
 of the neural systems is the associative recognition
 of complex structures. In this paper we deal with the
 ART networks developed by Carpenter and Grossberg
 {\cite{Carpenter87}}, originally
 as a model to explain the adaptive phenomena in visual systems.
 The family of ART algorithms belongs to group of unsupervised
 learning algorithms based on the theory of {\it adaptive resonance}
 motivated by the need to construct network sufficiently flexible
 to novel inputs and preserving previously learned patterns.
 ART model represents simple clustering algorithm, which
 has been complemented with the ability to generate
 new neurons if necessary. This is done by using a so called
 {\it vigilance parameter}. For the summary of the key results
 and examples of ART network applications see e.g.\cite{Honavar94}

These properties led us to the opinion that ART network
should be powerful to classify and compress
the magnetic intra-dot configurations
generated during the reversal.
First, we must encode the magnetic structure into
network input format.
We used encoding, where each dot magnetization field is replaced
by $N_{\rm c}$ effective magnetic moments located at interaction centers.
The moments income into $2 N_{\rm c}-$ dimensional vectors
{\cite{Horvath2001}} named as $\tilde{m}_i$, where $i$ is the dot label.
The vectors $\tilde{m}_i$ represent inputs of ART which compresses this
information into $N_{\rm w}$ output neurons named $\tilde{w}_j$.
Their structure is
\begin{eqnarray}
  \tilde{m}_i &\equiv& [ {\bf{m}}_{i 1}, {\bf{m}}_{i 2}, \ldots, {\bf{m}}_{i N_{\rm{c}}} ] \nonumber\\
 \tilde{w}_j &\equiv& [ {\bf{w}}_{j 1}, {\bf{w}}_{j 2},\ldots, {\bf{w}}_{j N_{\rm{c}}}]\nonumber
\end{eqnarray}
where $i\in\{1,2,\ldots N_{\rm{d}}\}$, $j\in\{1,2,\ldots N_{\rm{w}}\}$
and
\begin{equation}
 {\bf{m}}_{i n} = \frac{1}{N_{\rm{S}}} \sum_{l \in \Box_{i n}} {\bf{S}}_l
 \, ,
\label{Eqmin}
\end{equation}
index $n$ enumerates the spin blocks; ${\bf m}_{i n}$
represents locally averaged
dot microstate over the $N_{\rm{S}}$ spins (see Fig.~\ref{schema_m})
belonging to the square
element $\Box_{i n}$ of $i$th dot.
The match between ART input
$\tilde{m}_i$ and output $\tilde{w}_j$
is measured by the "magnetic" Euclidean distance
\begin{equation}
  \|{\tilde{w}_j} - \tilde{m}_i \| =
\sqrt{\sum_{n = 1}^{N_{\rm{c}}} (
  {\bf{w}}_{j n} - {\bf{m}}_{i n} ) \cdot ( {\bf{w}}_{j n} -
  {\bf{m}}_{i n} )^{\rm{T}}}
\label{Eucdis}
\end{equation}
($\rm{T}$ is superscript denotes the transposition).
\begin{figurehere}
\begin{center}
\epsfig{file=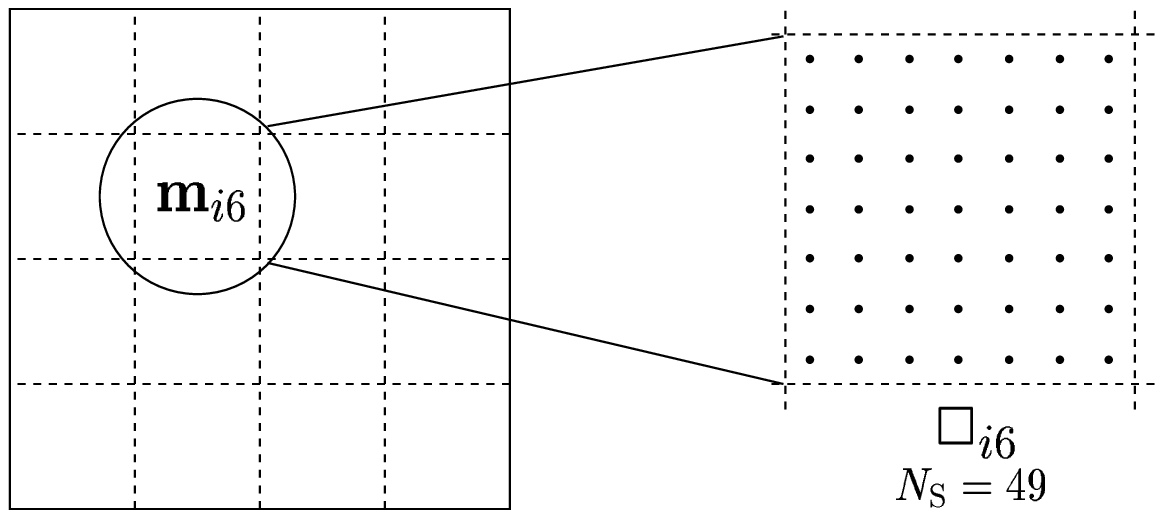,width=7.5cm,angle=0.0}
\caption{\small
 \label{schema_m}
The example of the averaging from Eq.(\ref{Eqmin})
showing the sixth square segment
$\Box_{i6}$ of $i$th dot including $N_{\rm S}=49$ spins.}
 \end{center}
\end{figurehere}
The learning algorithm is described in the following points
\begin{enumerate}
  \item[1.] {\it Initialization} setting $N_{\rm{w}} = 1$, $\tilde{w}_1
  = \tilde{m}_i$ for random selection $i \in \{ 1, 2, \ldots
  N_{\rm{d}} \}$.

  \item[2.] {\it Loop} through
 the set of the intra-dot magnetic patterns.
 For each randomly selected dot $i$ follows:
  \item[2.1] {\it Presentation} $\tilde{m}_i$ to the
    network. The learning is repeated for iteration steps
    $k = 1, 2, \ldots$

    \item[2.2] {\it Computing} of the actual
   index $j^{\ast} ( k, i )$ of the {\it winning neuron}
   $\tilde{w}_{j^{\ast}(k,i)}(k)$ trained by ${\tilde m}_i$
   according to the competitive rule
    \begin{equation}
      \begin{array}{rcl}
       j^{\ast} ( k, i ) = & \mbox{arg\, min} &  {\| {\tilde{w}_j} ( k ) -\tilde{m}_i \|} \\
       & ^{j = 1, 2, \ldots, N_{\rm{w}}} &
     \end{array}
    \end{equation}

    \item[2.3] {\it Comparing} $\|{\tilde{w}_{
     j^{\ast}(k,i)}} ( k ) -
    \tilde{m}_i \|$ to the
     vigilance parameter $\rho$.

      \item[2.3.1] {\it Update} of the weights
      via the Hebbian
      learning rule \cite{Haykin98}
      \begin{multline}
       \tilde{w}_{j^{\ast}(k,i)} ( k + 1 ) =
       \tilde{w}_{j^{\ast}(k,i)} ( k ) +
      \\
       \eta ( k ) \,
      \left( \tilde{m}_i - \tilde{w}_{j^{\ast}(k,i)} ( k ) \right) \, ,
      \end{multline}
      is applied if $\|{\tilde{w}_{j^{\ast}(k,i)}} - \tilde{m}_i \| \leq \rho$
      with the $k$-dependent learning rate parameter
      $\eta ( k ) = \eta_0
       \exp \left(- k / \tau_{\rm{learn}} \right)$.

      \item[2.3.2] {\it Creation} of the new neuron
      $\tilde{w}_{N_{\rm{w}} + 1}
      \leftarrow \tilde{m}_i$, $N_{\rm{w}}
      \leftarrow N_{\rm{w}} +
      1$ if $\|{\tilde{w}_{j^{\ast}(k,i)}} ( k ) - \tilde{m}_i \| > \rho$.

    \item[2.4] {\it Annihilation}
     of the neuron pair
     $[\,{\tilde w}_{z_1^{\ast}}(k),
     {\tilde w}_{z_2^{\ast}}(k)\,]$,
    \, $z_1^{\ast}< z_2^{\ast}$
     selected according to relation
    \begin{equation}
     \begin{array}{rclcc}
     [ z_1^{\ast}, z_2^{\ast} ] =
     & \mbox{arg \,min}
     & {\|{\tilde{w}_{z_1}}-\tilde{w}_{z_2} \|}
     & \quad
     &
     \\
     & ^{z_1, z_2} & &
     \end{array}
    \end{equation}
    if $\|{\tilde{w}_{
    z_1^{\ast}}}-\tilde{w}_{z_2^{\ast}} \|<\rho\,$.

    The product of annihilation
    $[ z_1^{\ast}, z_2^{\ast} ] \rightarrow z_1^{\ast}$,
    $N_{\rm{w}} \leftarrow  N_{\rm{w}} - 1$
    is the neuron
    $\tilde{w}_{z_1^{\ast}}$
    determined by the midpoint rule
    $\tilde{w}_{z_1^{\ast}}
    \leftarrow \frac{1}{2} ( \tilde{w}_{z_1^{\ast}} +
    \tilde{w}_{z_2^{\ast}} )$.
    After $\tilde{w}_{z_1^{\ast}}$ update ${\tilde w}_{j > z_2^{\ast}}$
    neurons undergo to {\rm collapse}:
   ${\tilde w}_{j-1}\leftarrow {\tilde w}_{j}$.

  \item[3.] {\it Stop criterion} is represented
  by the inequality
  \begin{equation}
    \frac{1}{N_{\rm{d}}}
   \sum_{i = 1}^{N_{\rm{d}}}
    \|
    {\tilde{w}_{j^{\ast}( k , i )}}(k+1) -
    \tilde{w}_{j^{\ast} ( k , i )}(k) \| <
    \varepsilon\,,
  \end{equation}
  where $\varepsilon$
  is small parameter. If the above inequality is
  not fulfilled the algorithm follows from the step 2
  with $k$ incremented by $1$.
  \end{enumerate}

The fulfillment of the last criterion means that network
attains a fixed point. By this way fixed neurons
$w_j,\,j=1,2, \ldots, N_{\rm w}$
represent the collection of the basic types
of the extracted intra-dot configurations.
In the case of magnetization reversal,
the ART network has been applied
independently to arrays at different time steps.
The typical intra-dot
and inter-dot remagnetization
transient states are presented
in the next section.

\vspace{0.5cm}
\hspace{-0.5cm}{\bf{4. RESULTS OF SIMULATION}}
\vspace{0.5cm}

The computations were performed for system of 100 dots
which each dot consists of 400 spins in external field
$H/J=-1$ and $K/J=0.1$ with initial
saturated state ${\bf M}(t=0)\cdot {\bf e}_x=1$, where
${\bf M}$ is the magnetization of array.

In the array simulations we observed the homogeneously
magnetized dots as well as dots showing the
nonuniform remagnetization. At the beginning of reversal
the opposite external magnetic field forms domain nucleation
centers at the dot corners.
The domains inclined by the external field
then grow and pass towards the dot centers
by diminishing the area of the opposite domains.
If the same polarity domains originating at different
parts (corners or edges) of the dot are separated
by the domain of opposite polarity,
the domain growth causes the formation
of two-fold domain wall - so called
soliton-antisoliton pairs \cite{Levy1999}.

The homogenization of intrinsic dot region is attained
by the driving effect of external magnetic field which
leads to wall motion annihilating
the soliton-antisoliton
pairs (see model Fig.~\ref{model_premag}).
The coherent rotation occurs only at the early stages
of the remagnetization for the sufficiently small dots when
the exchange length is sufficiently greater than size of dot.
This coherence locates nearly the corners of array.
The formation of the soliton-antisoliton domain walls is typical
for the central parts of the array and later stages
of the remagnetization.

\newpage

\begin{figurehere}
\begin{center}
\epsfig{file=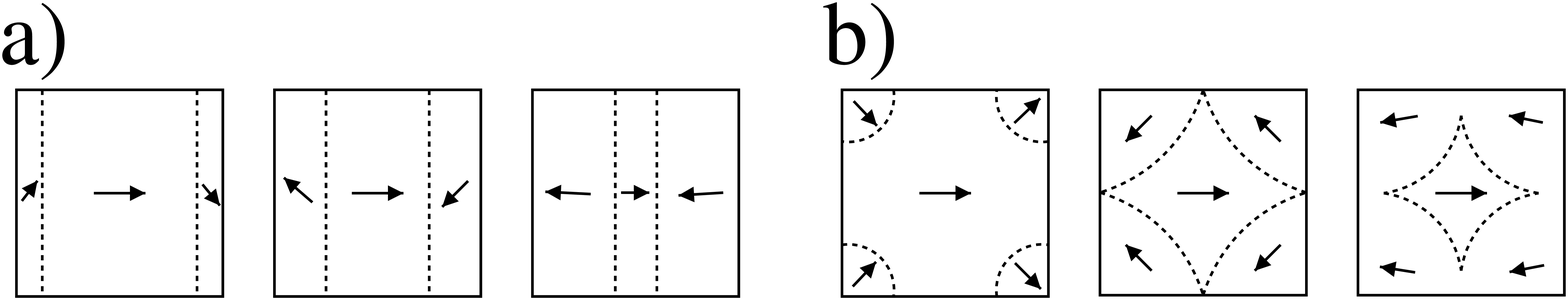,width=7.5cm,angle=0.0}
\caption{\small
\label{model_premag}
Two schemes of the principal transient states
of the nucleation, soliton-antisoliton pair formation,
domain motion and double-wall annihilation. The axial-symmetric
scenario with open-wall soliton-antisoliton pair (case~a)
is typical for the earlier transients
of remagnetization and non-centric dots,
whereas the center-symmetric soliton-antisoliton
closed-wall mode (case~b) is typical at the later
transient states and dots located near to the center
of array.}
\end{center}
\end{figurehere}

The main point of analysis of intra-dot magnetization
configurations deals with the implementation of ART network
treating the subsequent stages of the remagnetization.
The ART training was realized for the configurations
at times separated by the time step $\Delta t$ for
training parameters $\eta_0=0.1$,
$\tau_{\rm learn}=20$ and
$\varepsilon=10^{-6}$,
$\rho=1.15$.
The optimized choice of vigilance ($\rho=1.15$)
stems from the preliminary quasistatic calibration simulations
(see Fig.~\ref{vigNN}),
where vigilance was slowly decreased from the initial
large value $\rho=1.5$ (stabilizing one neuron) to the small
values pushing the network to the incorporation of many neurons.
The view point mediated by ART leads to the intra-dot taxonomy
depicted by Fig.~\ref{diagartNN}, where the nearest
neurons [in the sense of distance Eq.(\ref{Eucdis})]
belonging to different times are connected by arrows.
The ART model clearly distinguishes between
coherently rotating monodomain dots and
\begin{figurehere}
\begin{center}
\epsfig{file=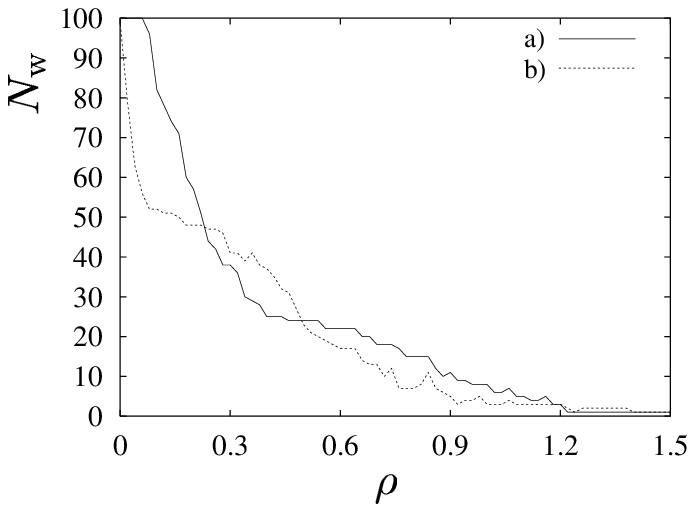,width=7.5cm,angle=0.0}
\caption{\small
\label{vigNN}
The pruning of ART network with the increasing vigilance parameter
for $K/J=0.1$ and $L_0=20$. The intra-dot configuration
is locally averaged over $N_{\rm S}=4$ spins [see Eq.(\ref{Eqmin})].
a)~the configuration near to the switching time ($t\simeq 850 \Delta t$);
b)~for the time when internal energy attains
   its maximum ($t\simeq 1100 \Delta t$).}
\end{center}
\end{figurehere}
dots including complex domains structures.
The 
increasing diversity of configurations occurring 
immediately after the switching time is 
self-adaptively reflected by the extended population of neurons.
The ART compression for later reversal
transient states is clearly caused by the small intra-dot
diversity within the array. The details can be studied by means
of the parameter $\rho$, which represents the virtual
"magnifying glass" capable to focus to the most
interesting final intra-dot transients
(see Fig.~\ref{detailNN} for $\rho=0.8$).
\end{multicols}

\vspace{0.5cm}

\begin{figurehere}
\begin{center}
\epsfig{file=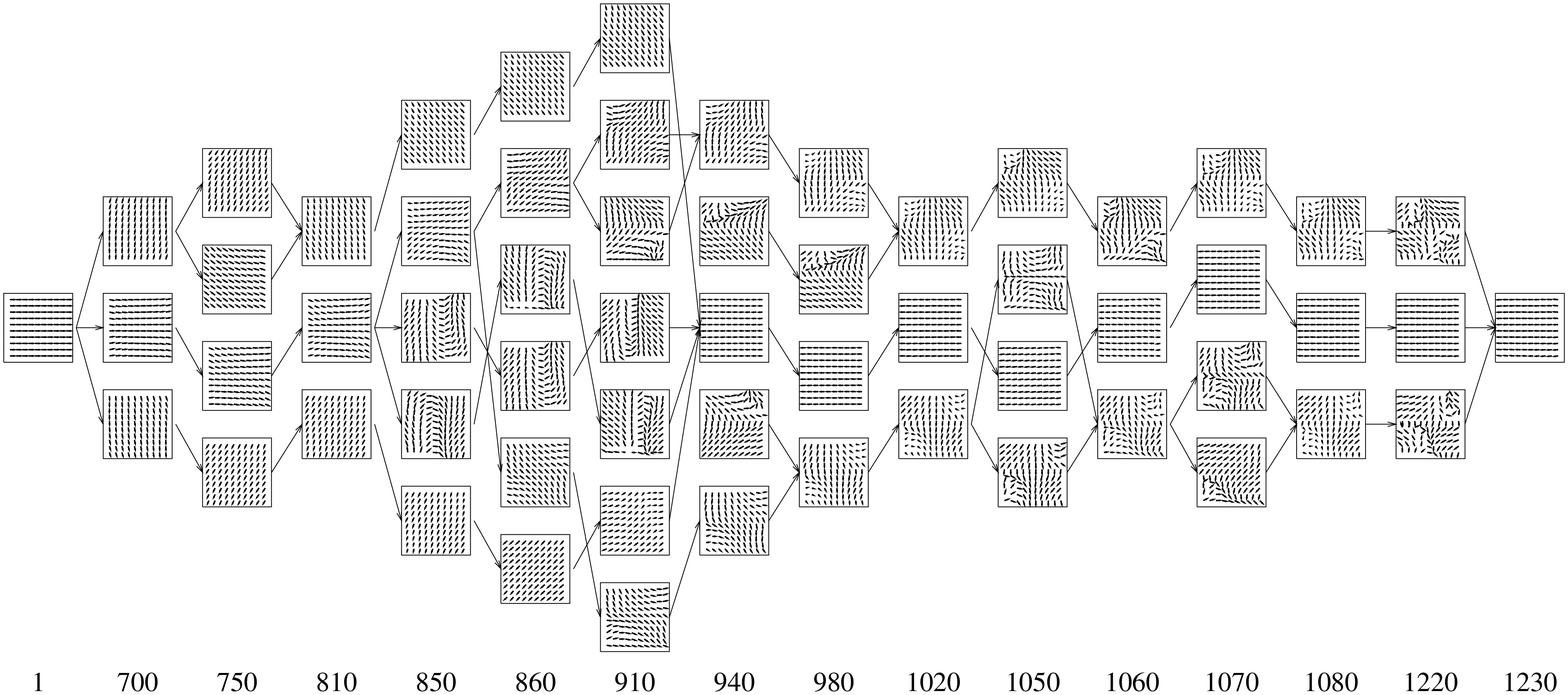,width=16.0cm,angle=0.0}
\caption{\small
\label{diagartNN}
The ART network representation of the magnetization reversal
for $K/J=0.1$, $L_0=20$.
The configuration belongs to $N_{\rm S}=4$ spin average
(see Eq.(\ref{Eqmin})). The arrows connect the closest
dots (in the sense of "magnetic" Euclidean distance from Eq.(\ref{Eucdis})).
The columns of intra-dot structures correspond to integer multiplies of
$\Delta t=0.01$ units.}
\end{center}
\end{figurehere}
\newpage
\begin{multicols}{2}
\begin{figurehere}
\begin{center}
\epsfig{file=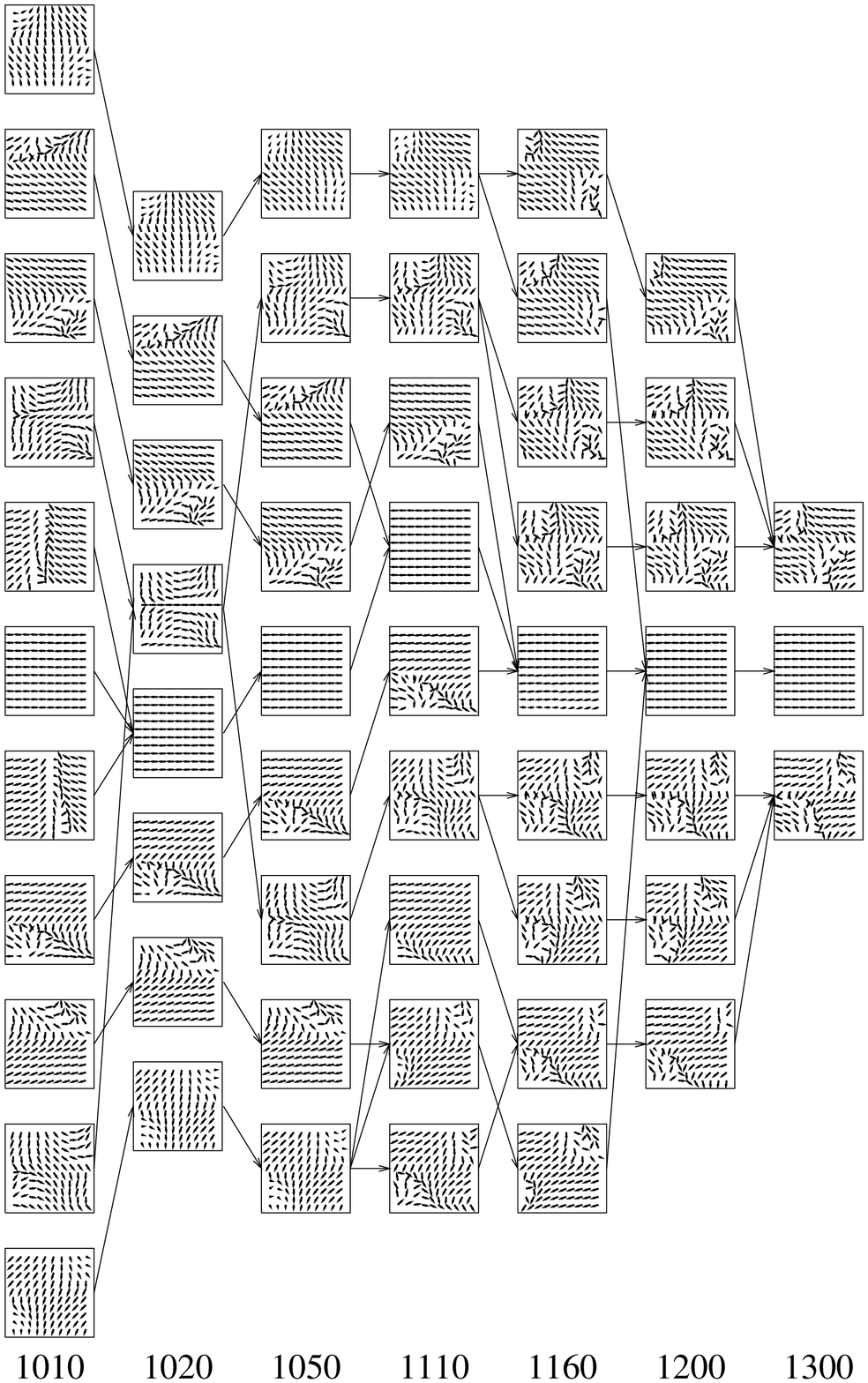,width=7.5cm,angle=0.0}
\caption{\small
\label{detailNN}
The detailed view on reversal in the ART representation for
sufficiently small $\rho=0.8$ showing the process of formation and annihilation
of intra-dot solitons (for the same parameters used in previous picture).}
\end{center}
\end{figurehere}

\vspace{0.2cm}
\hspace{-0.5cm}{\bf{5. CONCLUSIONS}}
\vspace{0.3cm}

The magnetization reversal has been simulated for
superlattice model of ultra-dense magnetic dot array.
Results of ART classification demonstrated that details
of reversal are determined by the interplay of exchange
and magnetostatic couplings, inter-dot domain wall
couplings and finite-size effects.
The consequence of periodic exchange-coupling cuts is
the formation of intra-dot soliton-antisoliton pairs.

The neural network is a new type of model which allow
studying of systems with many interacting pieces.
We have shown that ART diagrammatic viewpoint
is helpful in thinking how a dot switches
from one state to another within the large time
intervals, and how to extract the typical channels
of intra-dot evolution. 
The computational experience creates also believe
that advantages of ART paradigm should be valued
namely in 3D case, where visual classification
meets human recognition bounds.

\vspace{0.4cm}
\hspace{-0.5cm}{\bf{ACKNOWLEDGMENT}}
\vspace{0.2cm}

\hspace{-0.5cm}The work was supported by the
grant No. 1/6020/99 (Slovak Grant Agency).

\vspace{0.5cm}



\end{multicols}

\end{document}